\documentstyle[aps,tighten]{revtex}    
\def\bra{\langle} \def\ket{\rangle}    
\begin{document}    
\title{General Phase Matching Condition for Quantum Searching}    
\author{Gui-Lu Long$^{1,2,3,4}$, Li Xiao$^{1,2}$ and Yang Sun$^{5,1,2}$}    
\address{$^1$Department of Physics, Tsinghua University, Beijing 100084, P.R. China\\    
$^2$Key Laboratory for Quantum Information and Measurements, Ministry of Education, P.R.    
China\\    
$^3$Institute of Theoretical Physics, Chinese Academy of Sciences, Beijing, 100080, P.R.    
China\\    
$^4$Center of Atomic, Molecular and Nanosciences, Tsinghua University, Beijing 100084,    
P.R. China\\    
$^5$Department of Physics and Astronomy, University of Tennessee, Knoxville, TN 37996,    
U.S.A.}    
\date{\today}    
\maketitle    
    
\begin{abstract}    
We present a general phase matching condition for the quantum search algorithm with  
arbitrary unitary transformation and arbitrary phase rotations. We show by an explicit  
expression that the phase matching condition depends both on the unitary transformation  
$U$ and the initial state. Assuming that the initial amplitude distribution is an  
arbitrary superposition $\sin\theta_0|1\ket+\cos\theta_0 e^{i\delta}|2\ket$ with  
$|1\ket={1\over  \\sin\beta}\sum_{k}|\tau_k\ket\bra \tau_k|U|0\ket$ and   
$|2\ket={1\over \cos\beta}\sum_{i\ne \tau}|i\ket\bra i|U|0\ket$, where   
$|\tau_k\ket$ is a marked state and $\sin\beta=\sqrt{\sum_{k}|U_{\tau_k 0}|^2}$ is  
determined by the matrix elements of unitary transformation $U$ between $|\tau_k\ket$  
and the $|0\ket$ state, then   
the general phase matching condition    
is $\tan{\theta\over 2}\left[\cos 2\beta+\tan\theta_0\cos\delta\sin 2\beta\right]=    
\tan{\phi\over 2}\left[1-\tan\theta_0\sin\delta\sin 2\beta\tan{\theta \over 2}\right]$,    
where $\theta$ and $\phi$ are the phase rotation angles for $|0\ket$ and $|\tau_k\ket$, 
respectively.  
This generalizes previous conclusions in which the dependence   
of phase matching condition on $U$ and the initial state has been disguised.  
We show that several phase conditions    
previously discussed in the literature are special cases of this general one, which 
clarifies the question of which condition should be regarded as exact.  
   
\end{abstract}    
    
    
\section{Introduction}    
Grover's quantum search algorithm \cite{r1} is one of the most important developments in    
quantum computation. For searching a marked state in an unordered list, it achieves   
quadratic speedup over classical search algorithms.  In Grover's original paper  
\cite{r1}, each search step consists of two phase inversions and two Hadmard-Walsh  
transformations, and the initial state is an even distribution of the basis states.  
There have been several generalizations of the Grover algorithm. For instance, people 
have studied the cases with 
(1) more than one marked item \cite{r2};   
(2) an arbitrary unitary transformation instead of the Hadmard-Walsh transformation  
\cite{r2};   
(3) arbitrary initial distributions \cite{r3,r3p};  
(4) arbitrary phase rotations    
\cite{r5,r4}; and (5) arbitrarily entangled initial distribution  
\cite{carlini}.   
    
Arbitrary phase quantum searching has been extensively studied by our group. It was found  
that arbitrary  phase rotation of the marked state alone can not be used for a quantum  
search \cite{r5}.  It was later demonstrated \cite{r4} by an approximate treatment that a  
useful quantum  search algorithm can be constructed only if the  two phase rotations are  
equal, i.e. $\theta=\phi$ ($\theta$ and $\phi$ are the phase rotation angles for  
the $|0\ket$ state and the marked state, respectively).  
It is important that this phase matching    
condition should be satisfied during a searching process,  because the systematic error  
induced by phase mismatching is the dominant gate imperfection in the Grover algorithm  
\cite{r6}, and the error tolerance in phase mismatching is of the order $O(1/\sqrt{N})$. 
By the isomorphism between $SU(2)$ and $SO(3)$ group, an $SO(3)$    
picture for the quantum search algorithm has been established\cite{r7}. The advantage of  
this picture is that one can use simple geometrical method to treat quantum searching  
problems, even for cases where application of an analytical method is difficult. In this    
picture, a quantum search is described as a series of rotations in a 3-dimensional space.    
State  vector is represented by a polarization vector. The marked item corresponds to a  
point in the $z$-axis $(x,y,z)=(0,0,1)$ in space. The task of a quantum search is to  
rotate the polarization  
vector, initially lying near $(0,0,-1)$,    
to the target point $(0,0,1)$. During the searching process, the 3-dimension state vector 
(polarization vector) spans a cone in space, and the tip of the polarization vector draws
a circle in this cone. If  
the target point lies on this circle, the searching process can find the marked state.  
Using this $SO(3)$ picture, it  
was proven that the phase matching requirement $\theta=\phi$, which was obtained earlier  
through an approximation \cite{r4}, is an exact condition. Recently, this phase matching  
condition has been demonstrated in a 2-qubit system by the liquid NMR  
technique\cite{rexp}.     
  
Arbitrary phases have recently received much attention. Two papers have been published in  
Physical Review A, addressing particularly this issue\cite{r8,r3p}. In Ref. \cite{r8}, 
H{\o}yer  
discussed arbitrary phase rotations in quantum amplitude amplification, a generalization  
of Grover's quantum search algorithm. He obtained a phase condition $\tan{\phi\over   
2}=\tan{\theta\over 2}(1- 2a)$, where $a$ is the success probability of the search  
algorithm.  Using this phase condition, H{\o}yer constructed a quantum algorithm that  
searches a marked state with certainty. He also confirmed that the phase error tolerance  
is the order $O(1/\sqrt{N})$. By considering $\theta=\phi$ as an approximation to  
his phase condition, he can obtain our main results in Refs.\cite{r5,r4,r6,r7}. Since  
$a$ is of the order of $1/N$, the difference between  H{\o}yer condition and our condition  
$\theta=\phi$ is very small.  However, H{\o}yer claimed \cite{r8} that  $\tan{\phi\over  
2}=\tan{\theta\over 2}(1- 2a)$ is an exact phase condition and  $\theta=\phi$ is only an  
approximate one.  In another development, Biham {\it et al.} \cite{r3p} studied the 
arbitrary  
phase rotations in a quantum search algorithm that allows arbitrary phase rotations and  
arbitrary initial distribution using recursion relations. In their study, they  
found that in order for the algorithm to apply, the two rotation angles must be equal. 
The phase error tolerance in Ref. \cite{r3p} is found also to be the order 
$O(1/\sqrt{N})$.   
  
Although the main conclusions of these papers are similar, there is an apparent  
contradiction in the exact phase matching condition with arbitrary phases in a quantum  
search algorithm. In this paper, we will solve this paradox. More  
importantly, we have found a general phase matching condition for arbitrary phase  
rotations, with arbitrary unitary transformations and an initial distribution which is an  
arbitrary superposition of $|1\ket$ and $|2\ket$.   
We shall show that the paradox mentioned above can be solved by realizing a difference in  
the initial state distribution in the previous works. The two phase matching conditions  
are special cases of this general phase matching requirement. The phase matching  
requirement $\theta=\phi$ is obtained for a quantum    
search algorithm with an arbitrary unitary transformation $U$ and an initial    
distribution $U|0\ket$. The initial distribution of Grover's original algorithm and most  
of the generalizations of quantum search algorithm use this initial state.    Although  
H{\o}yer's initial state \cite{r8} also takes this form, the actual initial state for the  
searching, i.e. the process of repeated operation of the rotations, is {\it not},  
because he has to make some preparation to the initial state and this makes his initial  
state slightly different from $U|0\ket$. This makes  H{\o}yer's phase condition  
slightly different from ours. We shall also point out that other phase conditions are
special cases of the general phase matching condition derived in this paper.  
  
The paper is organized as follows. After this introduction, we briefly review the  
structure of a quantum search problem in section II. Here we particularly divide a quantum  
search algorithm into two parts: the quantum searching engine and the quantum database 
(the initial state).  In this way, one can see clearly the dependence of the phase matching  
condition on the unitary transformation $U$ and the initial distribution. This detailed  
dependence was ignored in previous discussions because the initial state has been taken as  
$U|0\ket$. In section III, we give the general phase matching condition using the $SO(3)$  
quantum searching picture. The advantage of this $SO(3)$ picture is the ease to treat  
quantum search problems in a simple geometrical picture. It is particularly useful in 
solving this problem. In section IV, we demonstrate our general phase matching conditions  
by several known examples. In section V, we discuss the influence of the phases on the  
computational complexity of the searching problem. Finally, a summary is  
given in section VI.   
   
\section{Structure of a quantum search algorithm}    
    
Let us review two basic aspects in a quantum search problem. First, one must have a    
searching  operation (we call it a search engine hereafter). Combining various  
generalizations, we can write a general quantum searching engine as the following  
operator:    
\begin{eqnarray}    
Q=-U I_{\gamma} U^{-1} I_{\tau},    
\label{engine}    
\end{eqnarray}    
where    
\begin{eqnarray}    
I_{\tau}&=&I+(e^{i \phi}-1)\sum_{k}|\tau_k\ket\bra \tau_k|,   
\nonumber\\    
I_{\gamma}&=&I+(e^{i\theta}-1)|\gamma\ket\bra \gamma| .    
\nonumber    
\end{eqnarray}    
Usually $|\gamma\ket$ is chosen as $|0\ket\equiv |0\cdots 0\ket$. Here $|\tau_k\ket$ is a  
marked state, and the summation runs over all the marked states. Thus,     
this quantum search engine can deal with cases with more than one marked state. We see    
that a quantum search engine is determined by the following factors: a unitary    
transformation $U$, two phase rotations and the marked states.    
    
Secondly, there must be a quantum database: the initial    
distribution $|\psi_0\ket$. This part is independent of the searching engine: for a given    
searching engine, the initial state may be prepared in various ways. However, a    
special form of the initial state makes the search problem simple. It was found    
in Refs.\cite{r9,r2} that the space span by $|1\ket$ and $|2\ket$ is invariant under the  
action of the quantum searching operator $Q$. If the  initial state is a superposition of  
these two state vectors, then the quantum search  problem can be dealt with in a  
2-dimensional space \cite{r2,r10}.   
  
In the literature, nearly all the initial distribution is chosen as   
\begin{eqnarray}  
U|0\ket=\sum_{i}|i\ket\bra i|U|0\ket=\sum_{k}|\tau_k\ket \bra \tau_k|U|0\ket+\sum_{i\ne  
\tau}|i\ket\bra i|U|0\ket=\sin\beta |1\ket+\cos\beta |2\ket,  
\end{eqnarray}  
where    
\begin{eqnarray}    
|1\ket &=&{1\over \sin\beta}\left(\sum_{k}|\tau_k\ket\bra\tau_k|\right)U|0\ket={1\over     
\sin\beta}\sum_{k}|\tau\ket U_{\tau_k\;0},\nonumber\\    
|2\ket &=&{1\over \cos\beta}\left(\sum_{i\ne \tau}|i\ket\bra i|\right)U|0\ket=    
{1\over \cos\beta}\sum_{i\ne \tau}|i\ket U_{i\;0},\nonumber\\    
\sin\beta &=&\sqrt{\sum_{k}|U_{\tau_k\;0}|^2}.\nonumber    
\end{eqnarray}    
 
For instance, in  the Grover algorithm, the evenly distributed initial state takes the  
form   
\begin{displaymath}    
|\psi_0\ket={1\over \sqrt{N}}\sum_{i=0}^{N-1}|i\ket    
={1 \over\sqrt{N}}|1\ket+\sqrt{N-1 \over N}|2 \ket.    
\end{displaymath}    
Of course, the form of the initial state may take a more general form. For instance,  
using the standard Grover searching engine where the unitary transformation is chosen as  
the Hadmard-Walsh transformation, the quantum search problem with arbitrary initial state  
was  studied in Ref. \cite{r3}. This was generalized to a quantum search  
engine with arbitrary phases and arbitrary unitary transformation in Ref. \cite{r3p}. In  
that case, the amplitudes of the marked states and unmarked states    
are  not tied together during a searching process, and one no longer has a    
2-dimensional rotation structure.     
    
In this paper, we restrict ourselves to the case where the initial state is an arbitrary     
superposition of $|1\ket$ and  $|2\ket$ (We refer this case as quasi-arbitrary initial  
distribution, to distinguish  this from that in Refs. \cite{r3,r3p}). The action of  
operator $Q$ on the two basis    
states are \cite{r4,r8,r12}    
\begin{eqnarray}    
Q\left[\begin{array}{c}|1\ket\\    
                       |2\ket    
	   \end{array}    
 \right]    
=\left[\begin{array}{cc}    
-e^{i\phi}(1+(e^{i\theta}-1)\sin^2\beta) & -(e^{i\theta}-1)\sqrt{\sin^2\beta  
(1-\sin^2\beta)} \\    
-e^{i\phi} (e^{i \theta}-1)\sin\beta\cos\beta &  
-e^{i\theta}+(e^{i\theta}-1)\sin^2\beta\end{array}    
\right]\left[\begin{array}{c}|1\ket\\    
                       |2\ket    
	   \end{array}    
 \right].    
\nonumber  
\end{eqnarray}    
 
Within this $U(2)$-formalism, after dropping a global phase factor, the initial state can    
be written most generally as     
\begin{eqnarray}    
|\psi_0\ket=\sin{\theta_0} |1\ket + \cos\theta_0 e^{i\delta}|2\ket.  
\label{ini}    
\end{eqnarray}    
 
\section{General phase matching condition}    
    
We now derive, in the $SO(3)$ picture,  
the phase matching requirement of the quantum searching engine    
(\ref{engine})    
with the initial state (\ref{ini}).    
The essence of the $SO(3)$ picture is that the quantum search operator in    
(\ref{engine}) is thought as a rotation operation in a 3-dimensional space.   
Explicitly, the matrix representing operator $Q$ in the basis span by $|1\ket$ and  
$|2\ket$ can be represented by a rotation in 3 dimensions,  
\begin{eqnarray}  
R_Q=\left[\begin{array}{ccc}R_{11} & R_{12} & R_{13}\\  
                            R_{21} & R_{22} & R_{23}\\  
	R_{31} & R_{32} & R_{33}\end{array}\right],  
\end{eqnarray}  
where   
\begin{eqnarray}  
R_{11}&=&\cos\phi (\cos^22\beta\cos\theta+\sin^2  
2\beta)+\cos2\beta\sin\theta\sin\phi),\nonumber\\  
R_{12}&=&\cos\phi\sin\theta\cos 2\beta-\cos\theta\sin\phi,\nonumber\\  
R_{13}&=&-\cos\phi \sin4\beta \sin^2{\theta \over 2}+\sin  
2\beta\sin\theta\sin\phi,\nonumber\\  
R_{21}&=&-\cos2\beta\cos\phi\sin\theta+\left(cos^2{\theta\over2}  
-\cos4\beta\sin^2{\theta\over 2}\right)\sin\phi,\nonumber\\  
R_{22}&=&\cos\theta\cos\phi+\cos 2\beta \sin\theta\sin\phi,\nonumber\\  
R_{23}&=&-\cos\phi\sin 2\beta\sin\theta-\sin 4\beta\sin^2{\theta\over  
2}\sin\phi,\nonumber\\  
R_{31}&=&-\sin 4\beta\sin^2{\theta \over 2},\nonumber\\  
R_{32}&=&\sin 2\beta\sin\theta,\nonumber\\  
R_{33}&=&\cos^2 2\beta+\cos\theta\sin^22\beta.\nonumber  
\end{eqnarray}  
This $SO(3)$ transformation corresponds to a rotation  about an axis $\vec{l}$ through a  
rotation angle $\alpha$, which  can be expressed as     
\begin{eqnarray}    
\vec{l}&=&\left[\begin{array}{c}    
\cot{\phi \over 2} \\    
1\\    
-\cot 2\beta \cot{\phi\over 2}+\cot{\theta\over 2} \csc 2\beta\end{array}    
\right],   
\label{axis}\\    
\alpha&=&\arccos\left[{1\over 4}(\cos4\beta+3)\cos\theta\cos\phi+(\sin    
2\beta)^2\left({1\over    
2}\cos\phi-\sin^2{\theta\over 2}\right)+\cos 2\beta\sin\theta\sin\phi\right].    
\label{angle}    
\end{eqnarray}    
During a searching process, the state of a quantum computer in general is    
\begin{eqnarray}    
|\psi\ket=(a'+b i)|1\ket+(c+d i)|2\ket,    
\nonumber  
\end{eqnarray}    
where $a'$, $b$, $c$ and $d$ are real numbers, satisfying the normalization    
condition $a'^2+b^2+c^2+d^2=1$. This state vector in the 2-dimensional space is    
represented by the polarization vector in the 3-dimensional space as \cite{r7}    
\begin{eqnarray}    
\vec{r}=\bra    
\psi|\vec{\sigma}|\psi\ket=\left[\begin{array}{c}x\\ y\\ z\end{array}\right]=    
\left[\begin{array}{c}2(a'c+b d)\\ 2(-b c+a d)\\ a'^2+b^2-c^2-d^2\end{array}\right],    
\label{polar}    
\end{eqnarray}    
where $\vec{\sigma}$ are the Pauli matrices.    
The probability of finding the marked state is    
\begin{eqnarray}    
P=a'^2+b^2=(z+1)/2.    
\label{prob}    
\end{eqnarray}    
These expressions make the understanding of the searching process very easy. 
For example, when the state vector is 
the marked state, its polarization vector is $(0,0,1)$ and the probability, according 
to Eq. (\ref{prob}), is 1. For the initial state, the polarization is about $(0,0,-1)$ and the 
probability for finding the marked state is nearly zero. 
 
Each searching iteration is a rotation of the polarization vector through angle $\alpha$.    
After $j$ iterations, the total angle rotated is     
\begin{eqnarray}    
\omega=j\alpha,    
\nonumber  
\end{eqnarray}    
and the polarization vector is rotated to     
\begin{eqnarray}    
\vec{r}_j=\vec{r}_0\cos\omega+\vec{l}_n(\vec{l}_n\cdot\vec{r}_0)(1-    
\cos\omega)+(\vec{l}_n\otimes \vec{r}_0)\sin\omega,    
\label{rstate}    
\end{eqnarray}    
where $\cdot$ and $\otimes$ are the ordinary scalar product and vector product operations.    
The vector $\vec{l}_n$ is the axis vector (\ref{axis}) normalized to unity. Using Eqs.    
(\ref{rstate}) and (\ref{prob}), the probability for finding the marked state    
can be easily calculated.    
    
During a searching process,    
the trajectory of the polarization vector (\ref{polar}) forms a cone    
whose rotational axis is given by (\ref{axis}). Starting from an initial position    
$\vec{r}_0$, the    
displacement vector $\vec{r}-\vec{r}_0$ is always perpendicular to the rotational axis. If    
the quantum searching process can find the marked state, then the vector    
$\vec{r}_f=(0,0,1)^T$ ($T$ means transpose) must be on the trajectory, thus    
$(\vec{r}_f-\vec{r}_0)\cdot \vec{l}=0$. By putting the initial state (\ref{ini}) into this    
equation, we obtain the following phase matching condition    
\begin{eqnarray}    
\tan{\theta\over 2}\left[\cos 2\beta+\tan\theta_0\cos\delta\sin 2\beta\right]=    
\tan{\phi\over 2}\left[1-\tan\theta_0\sin\delta\sin 2\beta\tan{\theta \over 2}\right].    
\label{match}    
\end{eqnarray}    
This is the general phase matching condition for a successful quantum search of    
marked states.  This phase matching condition tells us that the rotational angles depend  
on both the unitary transformation through $\beta$ and on the initial distribution through  
$\theta_0$ and $\delta$. In previous discussions, the dependence of the phase matching 
condition on the initial state was ignored because the initial state was taken as  
$U|0\ket=\sin\beta|1\ket+\cos\beta|2\ket$. In H{\o}yer's work \cite{r8},  
the initial state is  
modified before the search, and this makes the initial state different from $U|0\ket$,  
implicating the dependence on the initial state.  It should be pointed out that this  
condition is a necessary  condition for searching with certainty,  but not a sufficient  
one. Even if this condition is met, the probability of finding marked states   
is not guaranteed to be 1. The  standard Grover algorithm is one example. In the Grover  
algorithm \cite{r1},  the probability of finding the marked state with optimal iterations    
is $\sin^2\left[(2j_{op}+1)\beta\right]$. As  $\beta=\arcsin{1\over \sqrt{N}}$ is fixed,  
$(2j_{op}+1)\beta $ may not be exactly ${\pi \over 2}$.

\section{Examples of Phase Matching Conditions}    
    
It has been seen that the phase matching condition depends both on the structure of the    
quantum  searching engine and on the initial state. Now we discuss four examples, and  
show that the  general phase matching condition (\ref{match}) is satisfied in all these  
cases. The differences among these four examples are in their initial states.    
    
\subsection{$|\psi_0\ket=\sin(\theta_{init})|1\ket+\cos(\theta_{init})e^{i    
u}|2\ket$}    
    
In Ref. \cite{r8}, although the starting state is $U|0\ket=\sin\beta|1\ket+\cos\beta  
|2\ket$, some preparations have to be made before searching.  
First, the following state is obtained through 8  
steps\cite{r12}:   
\begin{displaymath}    
|\psi_{init}\ket=\sin(\theta_{init})|1\ket+\cos(\theta_{init})|2\ket.    
\end{displaymath}    
Before the searching iteration starts, a phase rotation $e^{i u}$ is made for the unmarked    
state. This leaves the initial state for the quantum search engine of the form,    
\begin{eqnarray}    
|\psi_{0}\ket=\sin(\theta_{init})|1\ket+\cos(\theta_{init})e^{i u}|2\ket.    
\label{thetaini}    
\end{eqnarray}    
In Eq. (\ref{thetaini}), $\theta_{init}={\pi\over 2}-m \vartheta$, where    
$\vartheta=\arcsin(|\sin{\theta\over 2}\sin 2\beta|)$, and $m$ is an integer    
\begin{eqnarray}    
m= \rm{INT}\left[({\pi \over 2}-\beta)/\vartheta\right].    
\label{m}    
\end{eqnarray}    
Here, INT[ ] means taking the nearest     
integer part.   
  
Since $\theta_{init}$ depends numerically on the quantities involved, an    
analytic proof is difficult. It has been carefully checked that, by using the initial  
state of  (\ref{thetaini}),  $\phi$ determined by $\tan{\phi\over  2}=  
\tan{\theta\over 2}(1-2\sin^2\beta)$ fulfills the general phase matching condition   
(\ref{match}). A numerical example is given in the appendix.    
    
\subsection{$|\psi_0\ket=U|0\ket=\sin\beta|1\ket+\cos\beta|2\ket$}    
    
This is the initial state that the most quantum search algorithms have taken. In 
Ref.\cite{r4}, the initial state is 
\begin{displaymath}    
|\psi_0\ket=U|0\ket=\sin\beta|1\ket+\cos\beta |2\ket.    
\end{displaymath}    
Putting this initial state into Eq. (\ref{match}) and letting $\theta_0=\beta$ and     
$\delta=0$, we obtain    
\begin{eqnarray}    
\tan{\theta\over 2}(\cos2\beta+\tan 2\beta\sin\beta)=\tan{\phi\over 2}.    
\nonumber  
\end{eqnarray}    
Using the fact that $\cos 2\beta+\tan \beta \sin    
2\beta=\cos^2\beta-\sin^2\beta+2\sin^2\beta=1$,    
we get     
\begin{displaymath}    
\tan{\theta\over 2}=\tan{\phi \over 2}, \qquad    
\rm{or} \quad \theta=\phi.    
\end{displaymath}    
This is the result that was obtained approximately in Ref.\cite{r4}, and exactly in Ref.     
\cite{r7} from an $SO(3)$ picture.    
    
\subsection{$|\psi_0\ket$ used by Brassard {\it et al.} \protect\cite{r13}}   
  
In Ref. \cite{r13}, a procedure was proposed for obtaining the marked state with    
certainty.  The strategy is to run the search algorithm $m'=m-1$ ($m$ is    
given  in (\ref{m})) number of iterations with $\theta=\phi=\pi$.  At this stage, the  
state vector of the quantum computer is just one step short of the marked state:  
$|\psi_0\ket=\sin((2m'+1)\beta)|1\ket +\cos((2m'+1)\beta)|2\ket$.  
Afterwards, one does one more search  
with $\theta$    
and  $\phi$ determined from the following equation    
\begin{eqnarray}    
\cot\{(2[\tilde{m}]+1)\beta\}   
= e^{i\phi}\sin (2 \beta) \left[-\cos (2\beta) + i   
\cot{\theta\over    
2}\right]^{-1}.    
\label{more}    
\end{eqnarray}    
  
We now show that the $\theta$ and $\phi$ determined in this way satisfy the general phase    
matching  condition (\ref{match}). Eq. (\ref{more}) is equivalent to two equations, which  
are the real and the imaginary part, respectively,    
\begin{eqnarray}    
\cos\phi\tan\theta_0\sin2\beta&=&-\cos2\beta,\nonumber\\    
\sin\phi\tan\theta_0\sin2\beta&=&\cot{\theta\over 2}.\nonumber    
\end{eqnarray}     
Here we have introduced the notation $\theta_0=(2m'+1)\beta$. It is then straightforward    
to show     
\begin{displaymath}    
\tan{\phi\over 2}={1-\cos\phi \over \sin\phi}={{1-{-\cos 2\beta \over \tan\theta_0\sin    
2\beta}} \over {\cot{\theta\over 2} \over \tan\theta_0\sin 2\beta}}=\tan{\theta\over    
2}\left[\cos 2\beta+\tan\theta_0 \sin 2\beta\right].    
\end{displaymath}    
This is exactly the general phase matching condition (\ref{match}) with $\delta=0$. It    
should be  pointed out that Eq. (\ref{more}) is a necessary and sufficient condition for  
finding the marked state with certainty. It determines the two angles uniquely.     
  
\subsection{``Difficult search problem limit" of arbitrary initial distribution by Biham  
{\it et al.} \protect\cite{r3p}}  
  
We see from the above examples that the phase matching condition strongly depends on the 
initial state. Recently, using an arbitrary initial distribution, Biham {\it et al.} have studied 
the general quantum search algorithm with arbitrary phase rotations\cite{r3p}. In particular,  
they obtained the phase matching condition $\theta=\phi$ which is the same as the case  
with $|\psi_0\ket=U|0\ket$. It seems contradicting that the apparent initial state  
dependence is missing here. The reason for this is that the phase condition of Biham {\it 
et al.}  
is obtained by using the ``difficult search problem limit": $N\gg N_\tau\geq 1$\cite{r3p},  
which gives the weighted averages $|\bar{k'}(0)|=O(W_k^{-1/2})$ and $|\bar{l'}(0)|=O(1)$.  
This is equivalent to  the case of $|\psi_0\ket=U|0\ket$. Thus it gives the same phase  
matching condition $\theta=\phi$. If this limit is not taken, then the phase matching  
condition can be varied greatly.

\section{The Computational complexity}  
  
Starting from the standard initial state $U|0\ket$ and the standard Grover's quantum  
search engine, the number of iterations is $O(\sqrt{N})$. If an quasi-arbitrary initial  
state is used instead of the standard initial state, the number of iterations will be  
different from $O(\sqrt{N})$. For instance, if the initial state is just the marked state,  
there is no need for search at all. If the initial state is the one after $m'$ iterations  
using the standard Grover as given in Ref. \cite{r13}, then one needs only one iteration.  
Using the $SO(3)$ picture of the quantum search, it is easy to study the computational 
complexity of the quantum search algorithm with arbitrary phases. Here, we present the 
results which can be proven through simple geometrical argument similar to the derivations 
given in Ref.\cite{rexact}: 
 
1) Given an initial state in Eq. (\ref{ini}) and an angle $\theta$,  
determining $\phi$ by solving Eq. (\ref{match}). (If the coefficient of the marked  
state is not real in the initial state, drop out a global phase factor in the initial  
state so that the coefficient of the marked state $|1\ket$ is real);  
 
2) Calculating the  
angle $\omega_{tot}$ between the initial state and the marked state in the $SO(3)$ picture 
by the following equation  
\begin{eqnarray}   
\omega_{tot}&=&\arccos\left({-K \sin(2\theta_0)(\cot({\phi\over   
2})\cos\delta+\sin\delta)-\cos2\theta_0\csc^2({\phi\over 2}) \over   
\sqrt{2K^2+1+\cot^2({\phi\over 2})+2 K   
(K\cos2\theta_0-\sin2\theta_0\sin\delta-\sin2\theta_0\cos\delta\cot({\phi\over  
2}))}}\right),  
\end{eqnarray}  
where  
\begin{eqnarray}   
K&=&\cot({\theta\over 2})\csc(2\beta)-\cot({\phi\over 2})\cot(2\beta),\\   
\cos\alpha&=&{1\over 4}(\cos(4\beta)+3)\cos\theta\cos\phi+\sin^2(2\beta)({1\over   
2}\cos\phi-\sin^2({\theta\over 2})+\cos2\beta\sin\theta\sin\phi));  
\end{eqnarray}   
 
3) Calculating the angle $\alpha$, which is the angle rotated by the quantum search engine  
in each iteration in the $SO(3)$ picture for given $\theta$ and the $\phi$ obtained  
through the phase matching condition.  
 
The number of iterations required to reach  
maximum probability in finding the marked state is given by   
\begin{eqnarray}  
j_{op}=\rm{INT}[{\omega_{tot} \over \alpha}].  
\end{eqnarray}  
Then maximum probability of finding the marked state is achieved by measuring the quantum  
computer at $j_{op}$ or $j_{op}+1$ step.   
  
To find the marked state with certainty, one has to modify the above procedure a little.  
If one wants to construct an quantum search engine that searches the marked state with  
certainty near a given $\theta$, one first uses the above procedure to obtain $j_{op}$.  
However, this quantum search engine does not guarantee to find the marked state with  
certainty. One has to use slightly different angles $\theta$ and $\phi$. They are  
determined by letting $\theta$ and $\phi$ as unknowns and solving simultaneously the phase matching  
condition (\ref{match}) and the equation $\omega/\alpha=J$ with $J>j_{op}$.  
Then the search algorithm with the angles so defined can find the marked state exactly 
when measured at the $J$th iteration. $J$ can be any number equal to or greater than $J_{op}$. 
A quantum search engine for finding the marked state with certainty with the standard  
initial state was recently given by Long in Ref.\cite{rexact}.   
  
\section{Summary}    
    
We have presented a general phase matching condition with arbitrary unitary    
transformations and an  arbitrary initial state superposed by $|1\ket$ and $|2\ket$.    
It has been shown that several phase conditions previously discussed in the literature are  
its special cases. Thus, there is a consistency between     
the results of \cite{r8} and \cite{r4} which have seemingly different expressions.   
The results in \cite{r13} and \cite{r3p} also satisfy this    
general phase  matching condition. The probability for obtaining the marked state has been  
given.    
    
This work is supported by the Major State Basic Research Development Program,     
Grant No. G200077400, the China National Natural Science Foundation Grant,     
No. 60073009, the Fok Ying Tung Education Foundation, and the     
Excellent Young University Teachers' Fund     
of Education Ministry of China. YS acknowledges also support from the advanced visiting    
scholar     
program of Tsinghua University, and the visiting scholar foundation of key laboratory of    
the    
Education Ministry of China.

\section*{appendix}    
    
We give a numerical proof that    
the phase condition in Ref.\cite{r8} is a special case of the general phase matching    
condition (\ref{match}). The initial state employed in Ref.\cite{r8} is     
\begin{displaymath}    
|\psi_0\ket=\sin(\theta_{init})|1\ket + \cos(\theta_{init})e^{i u}|2\ket,    
\end{displaymath}    
where $\theta_{init}={\pi\over 2}-m \vartheta$, $\vartheta=\sin{\theta\over    
2}\sin(2\beta)$, $\sin\beta=\sqrt{a}$, and $m=$INT$[({\pi \over    
2}-\beta)/\vartheta]+1$. $u$ is the difference of arguments $Q_{22}$ and    
$Q_{12}$, and $\phi=2\arctan \left[\tan{\theta \over 2} (1-2a)\right]$.    
    
Taking $a=2/400$, $\theta={\pi\over 2}$, we have     
\begin{eqnarray}    
\beta&=&\arcsin(\sqrt{a}),\nonumber\\     
\vartheta&=&\sin{\theta\over 2}\sin(2\beta),\nonumber\\    
m&=&16,\nonumber\\    
\phi&=&2\arctan\left[\tan{\theta\over 2} (1-2a)\right]=2\arctan{99\over 100};\nonumber\\    
Q_{11}&=&M_{22}=-{1\over 200}-{199 i\over 200},\nonumber\\    
Q_{12}&=&\left({1\over 200}-{i\over 200}\right)\sqrt{199},\nonumber\\    
Q_{21}&=&\left({1\over 200}-{i\over 200}\right)\sqrt{199} e^{2i\arctan{99\over    
100}}, \nonumber\\    
u&=& -(Arg[Q_{12}] - Arg[Q_{22}]),\nonumber\\    
\theta_{init}&=&{\pi \over 2}- m \vartheta. \nonumber    
\end{eqnarray}    
Putting the quantities $\theta$, $\phi$, $\beta$, $\delta=u$, $\theta_0=\theta_{init}$    
into Eq. (\ref{match}), we perform the calculation in {\it Mathematica}.     
With the number of digits up to    
150,    
the result for the left side of Eq. (\ref{match}) is 0.98723452878674 5048789300921936170    
7162274151777317 884870759108154687027972247    
696313773051896 66308976465471553486 4615871040404572923 23594964054244391216, and the one  
for the    
right hand side is exactly the same.     
   

\begin{thebibliography}{99}    
\bibitem{r1} L.K. Grover, {\it Phys. Rev. Lett.} {\bf 79}, 325 (1997).   
\bibitem{r2} L.K. Grover, {\it Phys. Rev. Lett.} {\bf 80}, 4329 (1998).   
\bibitem{r3} D. Biron, O. Biham, E. Biham, M. Grassl, and D.A. Lidar,    
 {\it Lecture Notes in Computer Science} vol. 1509, 140 - 147    
(Springer, 1998).   
\bibitem{r3p} E. Biham, O. Biham, D. Biron, M. Grassl, D.A. Lidar, and D. Shapira, {\it    
Phys. Rev.} {\bf A 63}, 012310 (2001).   
\bibitem{r5} G.L. Long, W.L. Zhang, Y.S. Li, and L. Niu,    
 {\it Commun. Theor. Phys.} {\bf 32}, 335 (1999).   
\bibitem{r4} G.L. Long, Y.S. Li, W.L. Zhang, and L. Niu, {\it Phys. Lett.} {\bf A 262}, 27    
(1999).    
\bibitem{carlini} A. Carlini and A. Hosoya, Phys. Lett. {\bf A 280}, 114 (2001).  
\bibitem{r6} G.L. Long, Y.S. Li, W.L. Zhang, and C.C. Tu,    
 {\it Phys. Rev} {\bf A 61}, 042305 (2000).   
 \bibitem{r7} G.L. Long, C.C. Tu, Y.S. Li, W.L. Zhang, and H.Y. Yan,    
 {\it J. Phys.} {\bf A 34}, 867 (2001). (See also    
 LANL eprint: quant-ph/9911004 for the original version, as some references were dropped  
out in the published version due to criticisms  
of the referee.)    
\bibitem{rexp} G. L. Long, H. Y. Yan, Y. S. Li {\it et al.}, to appear in Phys. Lett. A. 
Also available in quant-ph/0009059.  
\bibitem{r8} P. H{\o}yer, {\it Phys. Rev.} {\bf A 62}, 052304 (2000).   
\bibitem{r9} M. Boyer, G. Brassard, P. H{\o}yer, and A. Tapp, {\it Fortschritte Der  
Physik},  
{\bf 46}, 494 (1998).   
\bibitem{r10} S.X. Yu and C.P. Sun, LANL eprint: quant-ph/9903075.   
\bibitem{r12} We notice that there are some minor errors in Ref. \protect\cite{r8}, 
although they do not change the conclusion. 1) There is an exchange of the diagonal  
matrix elements in Eq. (3) of Ref.\cite{r8}; 2) We have to make a transpose to Eq. (9) in  
Ref.\cite{r8}; 3) During the preparation stage of the initial state in Ref.\cite{r8}, step  
(7) should be: apply operator $Q(A,\chi, \phi, \varphi)$ a total number of $m$ times on  
the first register conditionally on the 3rd register holding a 0.  
\bibitem{r13} G. Brassard, P. H{\o}yer, M. Mosca, and A. Tapp,    
 LANL eprint: quant-ph/0005055.  
\bibitem{rexact} G. L. Long, Grover algorithm with zero theoretical failure rate, to  
appear in Phys. Rev. A. Also available in quant-ph/0106071.  
\end{thebibliography}
\end{document}